\documentclass[twocolumn,showpacs,preprintnumbers,amsmath,amssymb,prl,superscriptaddress]{revtex4}

\usepackage{graphicx}
\usepackage{dcolumn}
\usepackage{bm}

\begin{document}


\title{Muon Spin Relaxation Studies of Superconductivity \\ in a Crystalline Array of Weakly Coupled Metal Nanoparticles }

\author{D.~Bono}
\affiliation{%
Kamerlingh Onnes Laboratory, Leiden University, P.O. Box 9504,
2300RA Leiden, The Netherlands
}%

\author{A.~Schnepf}
\affiliation{ Institut f\"ur Anorganische Chemie, Universit\"at
Karlsruhe, 76128 Karlsruhe, Germany
}%

\author{J.~Hartig}
\affiliation{ Institut f\"ur Anorganische Chemie, Universit\"at
Karlsruhe, 76128 Karlsruhe, Germany
}%

\author{H.~Schn\"ockel}
\affiliation{ Institut f\"ur Anorganische Chemie, Universit\"at
Karlsruhe, 76128 Karlsruhe, Germany
}%

\author{G.J.~Nieuwenhuys}
\affiliation{%
Kamerlingh Onnes Laboratory, Leiden University, P.O. Box 9504,
2300RA Leiden, The Netherlands
}%

\author{A.~Amato}
\affiliation{%
Laboratory for Muon Spin Spectroscopy, Paul Scherrer Institut,
CH-5232 Villigen PSI, Switzerland
}%

\author{L.J.~de Jongh}%
\affiliation{%
Kamerlingh Onnes Laboratory, Leiden University, P.O. Box 9504,
2300RA Leiden, The Netherlands
}%

\date{\today}

\begin{abstract}

We report Muon Spin Relaxation studies in weak transverse fields
of the superconductivity in the metal cluster compound,
Ga$_{84}$[N(SiMe$_{3}$)$_{2}$]$_{20}$\--Li$_{6}$Br$_{2}$(thf)$_{20}\cdot$\-2toluene.
The temperature and field dependence of the muon spin relaxation
rate and Knight shift clearly evidence type II bulk
superconductivity below $T_{\text{c}}\approx7.8$ K, with
$B_{\text{c1}}\approx 0.06$~T, $B_{\text{c2}}\approx 0.26 $~T,
$\kappa\sim 2$ and weak flux pinning. The data are well described
by the $s$-wave BCS model with weak electron-phonon coupling in
the clean limit. A qualitative explanation for the conduction
mechanism in this novel type of narrow band superconductor is
presented.
\end{abstract}


\maketitle

The chemical synthesis of molecular metal cluster compounds
presents an attractive bottom-up route for the generation of
self-organized nanostructures composed of 3D ordered arrays of
identical metal nanoparticles embedded in a dielectric matrix.
Until recently, such cluster solids were always found electrically
insulating. On the other hand, the strong similarity with
(super)conducting molecular crystals, as the alkali-doped
fullerenes (C$_{60}$), suggests that in principle metal cluster
compounds could also display  metallic conductivity (and even
superconductivity) due to intermolecular charge transfer. We have
indeed recently obtained compelling evidence from $^{69,71}$Ga-NMR
\cite{Bakharev06} and magnetization measurements for the
occurrence of \emph{band-type conductivity} in crystalline ordered
Ga$_{84}$ cluster compounds, composed of arrays of giant Ga$_{84}$
cluster molecules that display mixed-valence properties. In
addition, \emph{bulk type II superconductivity} was observed below
a transition temperature $T_{\text{c}}\approx 7.5$~K, much higher
in fact than known for bulk $\alpha$-Ga metal
($T_{\text{c}}\approx 1$~K). This material may thus represent a
first experimental realization of a theoretical model advanced by
Friedel in 1992 \cite{Friedel92}, who predicted that for a
\emph{crystalline array of identical metal nanoparticles}, a very
weak interparticle charge transfer can still yield
superconductivity with a relatively high $T_{\text{c}}$ value.

From the NMR study, a number of results were obtained. First, the
field and temperature dependence of the Knight shift can be well
fitted to BCS theory for weak electron-phonon coupling, a rather
surprising result since strong correlations would be expected for
such narrow band conductors as the present. Second, a very strong
sample dependence of the second critical field $B_{\text{c2}}$ was
revealed. As compared to $B_{\text{c2}}\approx 13.8$~T reported
previously \cite{Hagel02}, we found that for most of the samples
the minimum field of about 2~T required for the NMR signal (due to
signal-to-noise ratio) was already sufficient to suppress the
superconductivity completely. Only in a few samples
superconductivity could be detected by NMR up to about 5~T. By
contrast, magnetization studies showed diamagnetic signals for all
samples below similar values of $T_{\text{c}}=$7-8~K, but with
apparent $B_{\text{c2}}$ values as low as 0.3~T. Although an
explanation of this variation (by a factor 50!) in terms of
lattice defects
---~associated with orientational disorder in the molecular
crystal lattice~--- was proposed, definite proof for the
occurrence of bulk superconductivity in the low $B_{\text{c2}}$
samples seemed highly desirable. In order to check this
assumption, muon-Spin-Relaxation ($\mu$SR) experiments in small
transverse fields ($B_{\text{TF}}$) have therefore been performed.
As discussed below, they provide unambiguous proof for bulk type
II superconductivity, with a Ginzburg-Landau (GL) parameter
$\kappa\sim 2$, and are likewise in agreement with weak-coupling
BCS model predictions. A qualitative discussion of the nature of
the conduction mechanism in terms of intermolecular charge
fluctuations is presented at the end of the paper.

The crystal structure of the giant molecular cluster compounds
Ga$_{84}$, which is short for
Ga$_{84}$[N(SiMe$_{3}$)$_{2}$]$_{20}$\--Li$_{6}$Br$_{2}$(thf)$_{20}\cdot$\-2toluene,
is fully described in \cite{Schnepf01}. The $\mu$SR experiments
were performed on the sample labelled S3 in the previous work
\cite{Bakharev06}, which showed an excellent crystallinity in
x-ray experiments \cite{Schnepf01}, with typical dimensions of the
crystallites of 0.1~mm. Magnetization measurements in a commercial
SQUID magnetometer evidence a SC transition at $T_{\text{c}}=
7.75(5)$~K in zero external field. A $T=0$ extrapolation yields an
upper critical field $B_{\text{c2}}\approx 0.26$~T. Using the
expression $B_{\text{c2}}\approx\Phi_{0}/2\pi\xi^{2}$, where
$\Phi_{0}$ is the flux quantum, gives a coherence length $\xi\sim
35$~nm much larger than the inter-cluster distance, of order 2~nm.
The extrapolated thermodynamical critical field at $T=0$ was found
as $B_{\text{c1}}\approx 60$~mT. Using the general expression
given by Brandt for an ideal triangular GL vortex lattice,
relating $B_{\text{c2}}$, $B_{\text{c1}}$ and the GL parameter
$\kappa$ \cite{Brandt03}, we find $\kappa\sim 2$ and a London
penetration depth $\lambda\sim 70$~nm. This sample, which is
consequently close to the ``clean'' SC limit, indeed displays the
largest conducting fraction reached so far in a Ga$_{84}$ sample
(90\%), as found in the $^{69,71}$Ga NMR experiments performed on
several batches \cite{Bakharev06}.

Transverse field $\mu$SR is a powerful technique to probe type II
superconductivity locally \cite{Sonier00}, and can be performed in
any arbitrary field close to zero, contrary to NMR. By implanting
100\% spin polarized muons ($\mu^{+}$) in the material, the local
field they experience can be measured through their decay into,
among others, a positron, in a typical time window of
0.1-10~$\mu$s. We performed our experiments on the $\pi$M3 beam
line on the GPS spectrometer in PSI, Villigen, Switzerland, in the
temperature range $2~\text{K}\leq T\leq 10~\text{K}$, using
magnetic fields $0\leq B_{\text{TF}}\leq 0.4~$T. The Ga$_{84}$
sample (30~mg) is extremely air sensitive and must be kept in a
toluene solution with a sample/toluene mass ratio of order
$80/20$($\pm$10) to avoid loss of crystal solvent. We therefore
sealed it in an Al sample holder \footnote{The Al used is
comparable to the ``pure'' one of \cite{Kehr82}.}, using an Al
thickness of 0.35~mm between the incident $\mu^{+}$ beam and the
sample as a moderator for the muons. From the geometry our sample,
a pellet of thickness 0.55~mm and diameter 7~mm, and the
calibrated beam distribution, the $\mu^{+}$ fraction stopped in
the Al sample holder (plus the non conducting fraction of
Ga$_{84}$) called ``background'' in the following, is of order
25\%. This will appear as a temperature- and field-independent
signal in the whole temperature range. The $\mu^{+}$ fraction
stopped in toluene is of order 17\% and does not contribute to the
measured signal, as seen experimentally by a reduction of the
expected intensity by this amount. We attribute this reduction to
the formation of muonium in this solvent. Given the large fraction
of ``lost'' muons ($\sim 42$\%) we took data with good statistics,
with typically $40\times 10^{6}\ \mu^{+}$ per point.

      \begin{figure}[tbp!] \center
\includegraphics[width=1\linewidth]{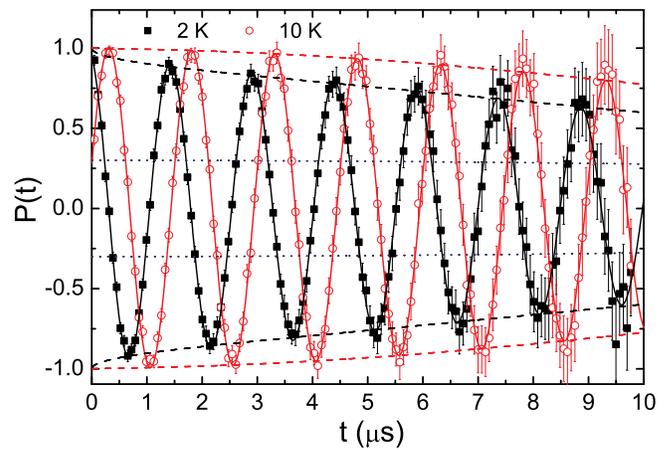}
      \caption{ \label{Asym} Polarization $P(t)$ of the $\mu^{+}$ spins,
      in a FC experiment ($H_{\text{TF}}=60~$mT). For clarity,
      only one of the two parts (imaginary or real)
      is presented for each $T$,
      in a rotating frame at a frequency $\gamma \times 55~$mT.
      The continuous lines are fits with Eq.~\ref{eqasym}.
      The dotted and dashed lines represent the envelope of the ``background''
      and the total signal, respectively.
      }
      \end{figure}

Figure~\ref{Asym} shows the time dependence of the $\mu^{+}$
polarization $P(t)$ along the axis defined by $B_{TF}$, for two
temperatures and $B_{\text{TF}}=60~\text{mT} \gtrsim
B_{\text{c1}}$, in a Field Cooled (FC) experiment. A clear
difference can be seen between the data measured at
$T=10~\text{K}>T_{\text{c}}$ and $T=2~\text{K}\ll T_{\text{c}}$.
Above $T_{\text{c}}$, the local field ($B_{\mu}$) distribution is
uniform in the sample and centered at the corresponding Larmor
frequency $\gamma B_{\mu}/2\pi$, where $\gamma$ is the $\mu^{+}$
gyromagnetic ratio. In real time, the relaxation of $P(t)$ is
therefore very slow. Below $T_{\text{c}}$, a flux-line lattice
(FLL) is created in the mixed state, resulting in a larger
distribution of $B_{\mu}$, i.e, a faster relaxation of $P(t)$.

\begin{figure}[tbp!] \center
\includegraphics[width=1\linewidth]{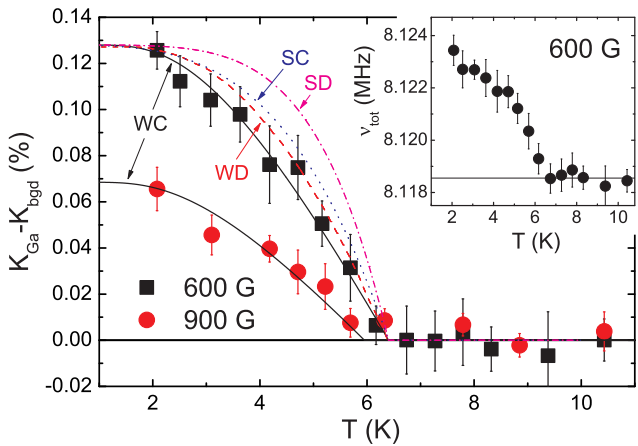}
      \caption{ \label{Shift} $T$-dependence of $K_{\text{Ga}}-K_{\text{bgd}}$,
      for two values of $B_{\text{TF}}$. Lines are BCS fits with a $s$-wave symmetry of the SC
      gap. ``W'' and ``S'' are for ``weak'' and ``strong'' electron-phonon
      coupling, ``C'' and ``D'' for ``clean'' and ``dirty'' limits.
      Insets: precession frequency $\nu_{\text{tot}}$ of the $\mu^{+}$ spins obtained with a single
      damped cosine fit, for $H_{\text{TF}}=60~$mT.
      The line shows the average value for $T>T_{\text{c}}$.
      }
      \end{figure}

In order to get a first \emph{qualitative} impression of the
behavior of the various relevant parameters, a single damped
cosine was fitted to the raw data. This model-independent analysis
already clearly shows a transition at $T_{\text{c}}\approx 6.4$
and 5.9~K, for $B_{TF}=60$ and 90~mT (corresponding to
$B_{\text{c2}}$ found in magnetization measurements at these
temperatures), in a FC experiment, in both the relaxation rate
(not shown) and the precession frequency, $\nu_{\text{tot}}$
[Fig~\ref{Shift}(inset)]. In a second step, the background
contribution has to be evaluated to get more \emph{quantitative}
information. To fit the $\mu^{+}$ polarization, we used the
phenomenological function
\begin{eqnarray}
\label{eqasym} P(t)&=&x
\exp(-[\sigma_{\text{Ga}}t]^{\alpha}-\sigma_{\text{nd}}^{2}t^{2})\cos(\omega_{\text{Ga}}
t+\phi) \nonumber
\\ && + (1-x)\exp(-\sigma_{\text{bgd}}^{2}t^{2})\cos(\omega_{\text{bgd}}
t+\phi)\ .
\end{eqnarray}
The first term accounts for the Ga$_{84}$ signal. We point out
that these $\mu^{+}$ are likely stopped in the
Ga$_{84}$[N(SiMe$_{3}$)$_{2}$]$_{20}^{4-}$ clusters, the other
locations in the crystal corresponding to regions having a
positive charge. The second term represents the background. The
relaxation rates $\sigma_{\text{nd}}$ and $\sigma_{\text{bgd}}$
are due to the nuclear dipolar fields on the respective sites
\cite{Sonier00}. We first determined the $T$- and
$B_{\text{TF}}$-independent parameters by comparing all the data.
The fits yield a fraction $x\approx 0.7$ of $\mu^{+}$ stopping in
Ga$_{84}$, $\sigma_{\text{nd}}\approx 0.06~\mu$s$^{-1}$ and
$\sigma_{\text{bgd}}\approx 0.04~\mu$s$^{-1}$, which are typical
values for nuclear relaxation rates
\cite{Amato97,Sonier00,Kehr82}. Although a rough approximation, a
Gaussian ($\alpha=2$) is usually suited to fit the FLL field
distribution in powder samples \cite{Aeppli87,Brandt03}. This
implies that the second moment of the field distribution,
$\langle\Delta B_{\mu}^{2}\rangle^{1/2}$, related to the
penetration depth $\lambda$, is directly proportional to the
$\mu^{+}$ relaxation rate due to the FLL distribution
($\sigma_{\text{Ga}}$). In our case a Gaussian cannot account for
the decay of the first $P(t)$ oscillations at low-$T$
(Fig.~\ref{Asym}) and we find instead a $T$- and
$H_{\text{TF}}$-independent $\alpha\approx0.5$. This suggests a
more complex FLL distribution \cite{Brandt88} in this cluster
compound. A complication of this $\alpha$ value is that the
corresponding field distribution does not converge unless a cutoff
(which could not be measured within the experimental resolution)
is introduced in the spectrum. To proceed, we therefore assume the
proportionality $\sigma_{\text{Ga}}^{2}\propto\langle\Delta
B_{\mu}^{2}\rangle$ to be still obeyed. We are then finally left
with only two free parameters, $\sigma_{\text{Ga}}$ and
$\nu_{\text{Ga}}$.

The $T$-dependence of the average shift
$K_{\text{Ga}}=(\omega_{\text{Ga}}-\gamma B_{TF})/\gamma B_{TF}$,
is displayed in Fig.~\ref{Shift} with the background shift
$K_{\text{bgd}}$ subtracted. This (muon) \emph{Knight shift} is
usually the sum of a $T$- and $B_\text{TF}$-independent orbital
shift $K_{\text{orb}}$ and a spin part $K_{S}$, proportional to
the $s$-electron susceptibility [and to the electronic density of
state at the Fermi level, $D(E_{F})$]. In the normal state (n), we
find a $T$- and $B_{\text{TF}}$-independent value (as expected in
metals) of $K_{\text{Ga,n}}\sim K_{\text{bgd}}\sim -0.17$\%. The
negative sign is due to the hyperfine coupling of the $\mu^{+}$
with the carriers, which is known to be strongly sample dependent
\cite{Schenk81} \footnote{Notice that there is \emph{a priori} no
reason to find $K_{\text{Ga,n}}\sim K_{\text{bgd}}$, a
relationship which is clear, however, from the narrow resonance
line measured at high-$T$.}. The present value is comparable to
those found in several heavy fermion compounds
\cite{Heffner86,Feyerherm92}.

Below $T_{c}$, the average precession frequency (inset of
Fig.~\ref{Shift}) is seen to increase continuously down to
$T\rightarrow0$ for both values of $B_{\text{TF}}$. Since the
Knight shift is negative, this implies a decrease of its absolute
value. In the mixed state, a possible additional shift
($K_{\text{d}}$) due to the demagnetization and Lorentz fields
would reduce the local field at the muon site. Therefore, it
cannot account for an increase of $\nu_{\text{tot}}$ and hence the
variation of the shift can be attributed to a decrease of the spin
part $|K_{S}|$ due to spin-singlet-pairing of the quasiparticles
below $T_{\text{c}}$ \footnote{Since the local field at the muon
site is involved, the demagnetizing and Lorentz field would cancel
for perfectly spherical superconducting particles. For
approximately spherical crystallites, as in our powder sample, the
resulting correction is therefore quite small. Demagnetizing
corrections arising from the shape of the sample holder are
likewise small, since it was a very flat disk and the transverse
field was parallel to the platelet.}.

Figure~\ref{Shift} shows that $K_{S}(T)$ is well fitted by a BCS
model with $s$-wave symmetry of the SC gap and weak
electron-phonon coupling (i.e., with a SC gap
$\Delta_{0}\approx1.76 k_{\text{B}}T_{\text{c}}$)
\cite{BCS57,Yosida58}. The decrease of the SC carrier density due
to the increase of $B_{\text{TF}}$ is directly observed through a
non-zero shift at $T\rightarrow 0$. It is also clear that the
reduction of $|K_{S}|$ is larger in the ``low'' field measurement,
as expected from a larger fraction of Cooper pairs \cite{Zheng02}.
The strong electron-phonon coupling and the dirty limit cases (the
latter being unlikely from the magnetization data) \cite{Rammer88}
are also presented in Fig.~\ref{Shift}. Their steeper variation
below $T_{\text{c}}$ clearly does not agree with the data.

      \begin{figure}[tbp!] \center
\includegraphics[width=1\linewidth]{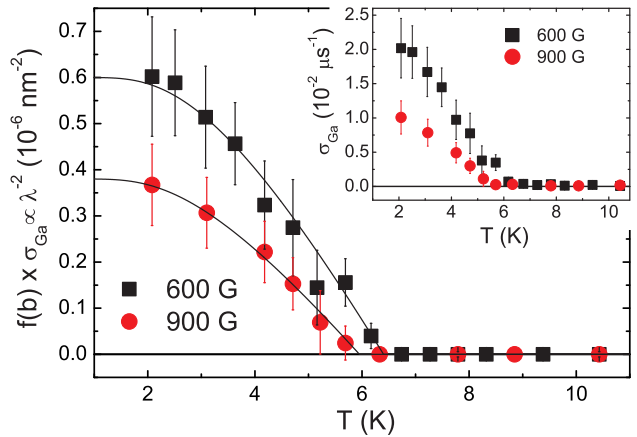}
      \caption{ \label{SigmadeT} $T$-dependence of $f(b)\sigma_{\text{Ga}}\propto \lambda^{-2}$.
      The lines are BCS fits (WC). Inset:
      $\sigma_{\text{Ga}}(T)$ from Eq.~\ref{eqasym}.
      }
      \end{figure}

The $T$-dependence of $\sigma_{\text{Ga}}$ is presented in the
inset of Fig.~\ref{SigmadeT}. The reduction of $T_{\text{c}}$ and
an increase of the internal field homogeneity (i.e., a decrease of
$\sigma_{\text{Ga}}$), is clearly observed when $B_{\text{TF}}$ is
increased. In the extensively studied cuprates and heavy fermions,
the conditions $\kappa \gg 1$ and $B _{\text{TF}}\ll
B_{\text{c2}}$ imply that the ratio
$\sigma_{\text{Ga}}/\lambda^{-2}$ is roughly $T$- and $B
_{\text{TF}}$-independent \cite{Amato97,Sonier00}. Here, the
artificial reduction of $\sigma_{\text{Ga}}$ due to the overlap of
the vortices near $B_{\text{c2}}$ has to be taken into account. In
the framework of the GL theory, Brandt derived the quantity
$\lambda^{2}\langle \Delta B_{\mu}^{2}\rangle^{1/2}$ as a function
of $\kappa$ and $b$, where $b$ is the average internal field
divided by $B_{\text{c2}}(T)$ \cite{Brandt03}. Calling this
quantity $f(b)$, the function $\lambda^{-2}\propto
f(b)\sigma_{\text{Ga}}$ is presented in Fig.~\ref{SigmadeT},
taking $\kappa= 2$. The qualitative variation of $\lambda^{-2}(T)$
is very similar to the $T$-dependence of $K_{S}$, as expected in
the BCS model in small fields \cite{BCS57,Yosida58}, and as shown
by the fit with weak electron-phonon coupling in the clean limit
(Fig.~\ref{SigmadeT}) \footnote{Quantitatively, the value
$1/\sqrt{f(b)\sigma_{\text{Ga}}(0)}\sim 1000$~nm is one order of
magnitude larger than $\lambda$ evaluated from the magnetization
data. This may be due to both the non-Gaussian fit and the non
ideal GL triangular FLL case.}.

      \begin{figure}[bp!] \center
\includegraphics[width=1\linewidth]{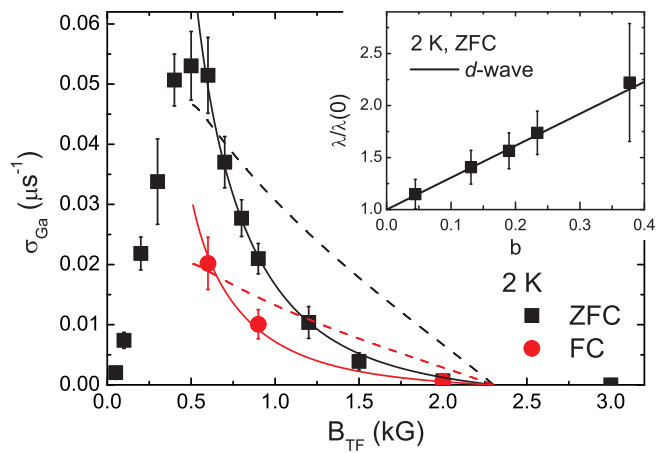}
      \caption{ \label{SigmadeH} $H_{\text{TF}}$-dependence of
      $\sigma_{\text{Ga}}$, in ZFC and FC conditions.
      The pinning accounts for the difference.
      Inset: $\lambda(b)/\lambda(0)$. The lines are computations
      with $\kappa=2$, a constant $\lambda$ ($s$-wave, dashed
      lines) and $\lambda/\lambda(0)\propto 1+3.1(5)b$ ($d$-wave, solid
      lines). We note that $b=0$ when $B_{TF}\leq B_{c1}$
      \cite{Brandt03}.
      }
      \end{figure}

We finally discuss the field dependence of $\sigma_{\text{Ga}}$ at
low-$T$. We measured $\sigma_{\text{Ga}}$ from the pure
diamagnetic state up to the normal state by increasing the field
from a Zero Field Cooled (ZFC) condition at 2~K, from 0 up to
$B_{TF}>B_{\text{c2}}$. The results of the fits with
Eq.~\ref{eqasym} are presented in Fig.~\ref{SigmadeH}. For small
fields ($B_{\text{TF}}\lesssim B_{\text{c1}}$),
$\sigma_{\text{Ga}}$ continuously increases with $B_{\text{TF}}$.
A strong inhomogeneity of the electron density of state in
Ga$_{84}$, as well as the powder nature of the sample as suggested
in \cite{Aeppli87}, may lead to some FLL bending around SC grains
and an inhomogeneity responsible for a non-zero
$\sigma_{\text{Ga}}$. When $B_{\text{TF}}\gtrsim B_{\text{c1}}$,
$\sigma_{\text{Ga}}$ reflects both the FLL organization
\cite{Brandt03} and the field dependence of $\lambda^{-2}$. For
high fields ($B_{\text{TF}}\gtrsim B_{\text{c2}}\sim 0.2$~T at
2~K), the superconductivity is destroyed and an homogeneous field
($\sigma_{Ga}=0$) is recovered. The ratio
$\lambda/\lambda(0)\propto [f(b)\sigma_{\text{Ga}}]^{-2}$ is
presented in Fig.~\ref{SigmadeH} as a function of $b$. In a BCS
model with $s$-wave symmetry, the penetration depth $\lambda$ is
expected to be weakly field-dependent \cite{Khasanov05}, with a
roughly linear variation of $\sigma_{Ga}$. However, a linear
variation of $\lambda$, quantitatively comparable to the type II
superconductor NbSe$_{2}$ \cite{Sonier00}, is measured and is
consistent with a $d$-wave symmetry of the SC gap (continuous
lines) \cite{Yip92,Sonier00}. Moreover, the predicted
$\sigma_{Ga}$ variation, with a constant $\lambda$, does not fit
the data (dashed line). The theoretical computations are usually
performed in an equilibrium FLL, which does not really correspond
to our case. Indeed, the pinning of the vortices increases the
local field distribution, as seen by the usual
\cite{Aeppli87,Le92} reduction of $\sigma_{\text{Ga}}$ in the FC
case (which also reduces the sensitivity of the experiment).
However, the shape of the variation does not seem to be changed.

Finally, we address the nature of the conduction mechanism in
Ga$_{84}$. In the Friedel model, referred to in the introduction,
conduction would occur in a narrow band originating from a
molecular energy level near $E_{F}$ of the Ga$_{84}$ cluster, with
a width proportional to the intercluster , i.e. anion-anion charge
transfer $t$. In view of the ligand shells around the Ga$_{84}$
cluster cores, $t$ is however expected to be quite small, of order
0.01~eV. In a Hubbard model approach, $t$ would have to compete
with the on-site Coulomb interaction (Hubbard-$U$) of the anion,
estimated at about 0.5~eV (three times smaller than for C$_{60}$).
In such a narrow-band system, strong electron-electron
correlations are expected whereas the NMR and $\mu$SR experiments
both indicate nearly free electron behavior (Korringa constants,
Knight shifts). It appears relevant, therefore, to consider other
possible charge fluctuations.

In analogy with the well-known models proposed for the transition
metal compounds \cite{Zaanen85}, one such possibility could be an
electron transfer from the cluster anion to the Li-cation, in
combination with conduction in the cation band derived from the
cation-cation overlap. Preliminary estimates, based on the cation
ionization energy, the anion electron affinity and the Madelung
energies involved, indicate the anion-cation charge transfer
process to be almost energetically neutral, i.e. could in fact be
quite small, smaller perhaps than the width of the cation band.
The latter is difficult to estimate but could be larger than $t$.
In this scenario the itinerant carriers would reside mainly in the
cation band, which could explain why in the measured Ga-NMR
spectra the different Ga sites in the cluster are clearly
resolved, as in an insulating Ga compound. Clearly, more
sophisticated calculations are needed to further elucidate the
conduction processes in this exciting novel type of molecular
(super)conductor.

The $\mu$SR data were collected at the S$\mu$S, PSI, Villigen,
Switzerland. This work is part of the research program of the
``Stichting FOM'' and is partially funded by the EC-RTN
``QuEMolNa'' (No. MRTN-CT-2003-504880), the EC-Network of
Excellence ``MAGMANet'' (No. 515767-2) (Leiden), the DFG-Centre of
Functional Nanostructures (Karlsruhe), and the European Commission
under the 6th Framework Programme through the Key Action:
Strengthening the European Research Area, Research Infrastructures
(No. RII3-CT-2004-505925).


\end{document}